\documentclass[fleqn,twoside]{article}

\usepackage{espcrc2}

\usepackage{graphicx}

\usepackage[figuresright]{rotating}

\newcommand{\AmS}{{\protect\the\textfont2

  A\kern-.1667em\lower.5ex\hbox{M}\kern-.125emS}}

\title{Exclusive Measurement of the $pp \to nn\pi^+\pi^+$ Reaction at 1.1 GeV}
 
\author{T.~Skorodko\address[PIT]{Physikalisches Institut der Universit\"at
  T\"ubingen, D-72076 T\"ubingen, Germany},
M.~Bashkanov\addressmark[PIT],
 D.~Bogoslawsky\address[JINR]{Joint Institute for Nuclear Research, Dubna,
  Russia},
H.~Cal\'en\address[SL]{The Svedberg Laboratory, Uppsala, Sweden},
H.~Clement\addressmark[PIT],
E.~Doroshkevich\addressmark[PIT],
L.~Demiroers\address[HU]{Hamburg University, Hamburg, Germany},
C.~Ekstr\"om\addressmark[SL],
K.~Fransson\addressmark[SL],
L.~Gustafsson\address[UU]{Uppsala University, Uppsala,Sweden},
B.~H\"oistad\addressmark[UU],
G.~Ivanov\addressmark[JINR],
M.~Jacewicz\addressmark[UU],
E.~Jiganov\addressmark[JINR],
T.~Johansson\addressmark[UU],
O.~Khakimova\addressmark[PIT],
S.~Keleta\addressmark[UU],
I.~Koch\addressmark[UU],
F.~Kren\addressmark[PIT],
S.~Kullander\addressmark[UU],
A.~Kup\'s\'c\addressmark[SL],
P.~Marciniewski\addressmark[SL],
R.~Meier\addressmark[PIT],
B.~Morosov\addressmark[JINR],
C.~Pauly\address[FJ]{Forschungszentrum J\"ulich, Germany},
H.~Petr{\'e}n.\addressmark[UU],
Y.~Petukhov\addressmark[JINR],
A.~Povtorejko\addressmark[JINR],
R.J.M.Y.~Ruber\addressmark[SL],
K.~Sch\"onning\addressmark[UU],
W.~Scobel\addressmark[HU],
B.~Shwartz\address[BINP]{Budker Institute of Nuclear Physics, Novosibirsk,
  Russia},
J.~Stepaniak\address[SINS]{Soltan Institute of Nuclear Studies, Warsaw and
  Lodz, Poland},
P.~Th\"orngren-Engblom\addressmark[UU],
V.~Tikhomirov\addressmark[JINR],
G.J.~Wagner\addressmark[PIT],
M.~Wolke\addressmark[UU],
A.~Yamamoto\address[HEARO]{High Energy Accelerator Research Organization,
  Tsukuba, Japan},
 J.~Zabierowski\addressmark[SINS],
and
J.~Zlomanczuk\addressmark[UU]}

\begin{document}

\begin{abstract}

First exclusive data for the $pp \to nn\pi^+\pi^+$ reaction have been
obtained at CELSIUS with the WASA detector setup at a beam energy  
of $T_p$ = 1.1 GeV. Total and differential cross sections disagree with
theoretical calculations, which predict the $\Delta\Delta$ excitation to be
the dominant process at this beam energy. Instead the data require the
excitation of a higher-lying $\Delta$ state, most likely the $\Delta(1600)$, to
be the leading process.
\vspace{1pc}

\end{abstract}


\maketitle

Two-pion production in nucleon-nucleon collisions connects $\pi\pi$ dynamics
with baryon and baryon-baryon degrees of freedom. Among the various reaction
channels the  $pp \to nn\pi^+\pi^+$ reaction is special, since the direct
excitation of $N^*$ resonances and their subsequent decay into the $\pi^+\pi^+$
channel is excluded by isospin. Hence it was expected that in the energy region
considered here only the $\Delta\Delta$ process would play the dominant
role. Indeed, the detailed calculations from the Valencia group \cite{alv}
predict the  $\Delta\Delta$ excitation to be the leading process at energies
$T_p >$ 1 GeV. However, in a recent isospin
decomposition of the total cross sections for the various $NN\pi\pi$ exit
channels we have shown \cite{tsi} that this assumption is inconsistent
with the experimental total cross sections. 

Due to the particular selectivity of the $pp \to nn\pi^+\pi^+$ reaction only
$I = 3/2$ single resonance excitations can contribute. Therefore we 
proposed \cite{tsi} that the excitation 
of a higher-lying $\Delta$ state, presumably the $\Delta(1600)$ might be the
leading process in this channel, since due to kinematics the $\Delta(1232)$
state cannot decay by emission of two pions.

A more recent theoretical study \cite{xu} of two-pion production in $NN$
collisions includes many additional processes, such as {\it e.g.}  nucleon pole
terms. That way good agreement is obtained with the
total cross section data for the $pp \to nn\pi^+\pi^+$ channel, however, at
the same time the $pp \to pp\pi^0\pi^0$ cross section is massively
overestimated for $T_p >$ 1 GeV.

In order to shed more light into this conflicting situation and since there
exist no differential cross sections at all for this channel we have
undertaken exclusive measurements of the $pp \to nn\pi^+\pi^+$ reaction at 
$T_p$ = 1.1 GeV using the WASA detector \cite{barg} with the hydrogen pellet 
target system at the CELSIUS storage 
ring of the Theodor Svedberg Laboratory in Uppsala. The detector has nearly
full angular coverage 
for the detection of charged and uncharged particles. 

The forward detector part 
consists of a thin-walled window plastic scintillator hodoscope (FWC) at the
exit of the scattering chamber, followed by straw tracker, plastic
scintillator quirl (FTH), forward range hodoscope (FRH) consisting of 24
cake-like segments and, finally, the forward interleaving (FRI) 
and veto hodoscopes.

The central detector comprises in its inner part a 
thin-walled superconducting magnet containing a minidrift chamber for tracking
and in its outer part a plastic scintillator barrel surrounded by an
electromagnetic calorimeter consisting of 1012 CsI (Na) crystals. The
positively charged pions were detected and identified in the central
detector.  

Neutrons were detected in the forward detector. They
were identified by the requirement of having no signal in the thin-walled
window hodoscope, straw tracker and quirl, however a signal due to recoil
protons in the range hodoscope. If the recoil protons were produced within the
first three layers of the FRH, then these recoil protons could also be
detected in the succeeding Forward Range Interleaving (FRI) hodoscope
\cite{HH}, which provides a more detailed angular information for the primary
neutrons.

The trigger was set to two charged particles in the central
detector and two neutron candidates in the forward detector. A neutron
candidate was identified by having an energy deposit (by recoil protons) of
larger than 40 MeV in a segment of the FRH with simultaneously zero hits
in the preceding thin-walled detectors FWC, straw tracker and FTH.  

The efficiency of the neutron detection was determined by means of the $pp \to
pn\pi^+$ reaction, which was identified by detecting $p$ and $\pi^+$ in the
central detector yielding a kinematically complete measurement.  From the
knowledge of the four-vectors for $p$ and $\pi^+$ the direction of the emitted
neutron could be reconstructed. Comparison of the expected hits in the
forward detector with actually identified neutron events
provides the desired information on the efficiency.

The efficiency and acceptance correction of the full data has
been performed with help of Monte Carlo simulations of the detector setup and
performance.
The absolute normalization of the data was achieved by normalizing to 
elastic scattering data --- measured simultaneously with the $pp \to
nn\pi^+\pi^+$ data.
 
With the four-vectors of the two pions and the
angular directions of the two neutrons we have measured the $pp \to
nn\pi^+\pi^+$ events with two kinematic overconstraints. Thus the data were
subjected to a corresponding kinematic fit. 

\begin{figure} [t]
\begin{center}

\includegraphics[width=0.49\textwidth]{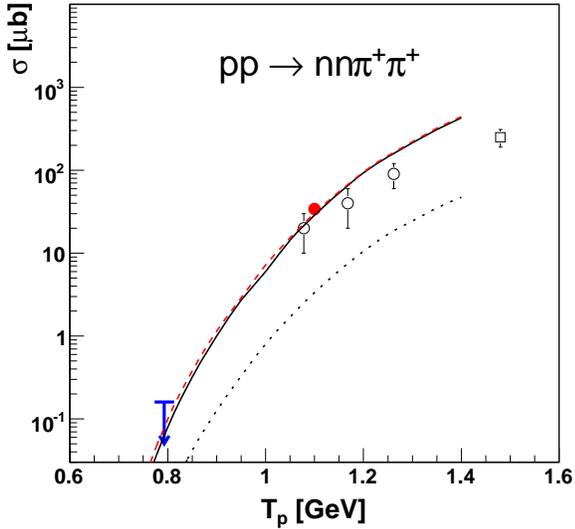}

\caption{ 
   Energy dependence of the total cross section for the $pp
   \to nn\pi^+\pi^+$ reaction. Open symbols denote previous bubble chamber
   measurements \cite{shim,eis}. The arrow gives the upper limit
   obtained at COSY-TOF \cite{evd} and the filled circle represents the
   experimental result of this work. The dotted line shows $\Delta\Delta$
   calculations as used 
   for the description of the $pp \to pp\pi^0\pi^0$ reaction \cite{deldel}. The
   dashed line represents calculations of the $\Delta(1600) \to \Delta \pi$
   process and the solid line gives the coherent sum of both processes. 
}
\label{fig1}
\end{center}
\end{figure}

\begin{figure}
\begin{center}

\includegraphics[width=0.23\textwidth]{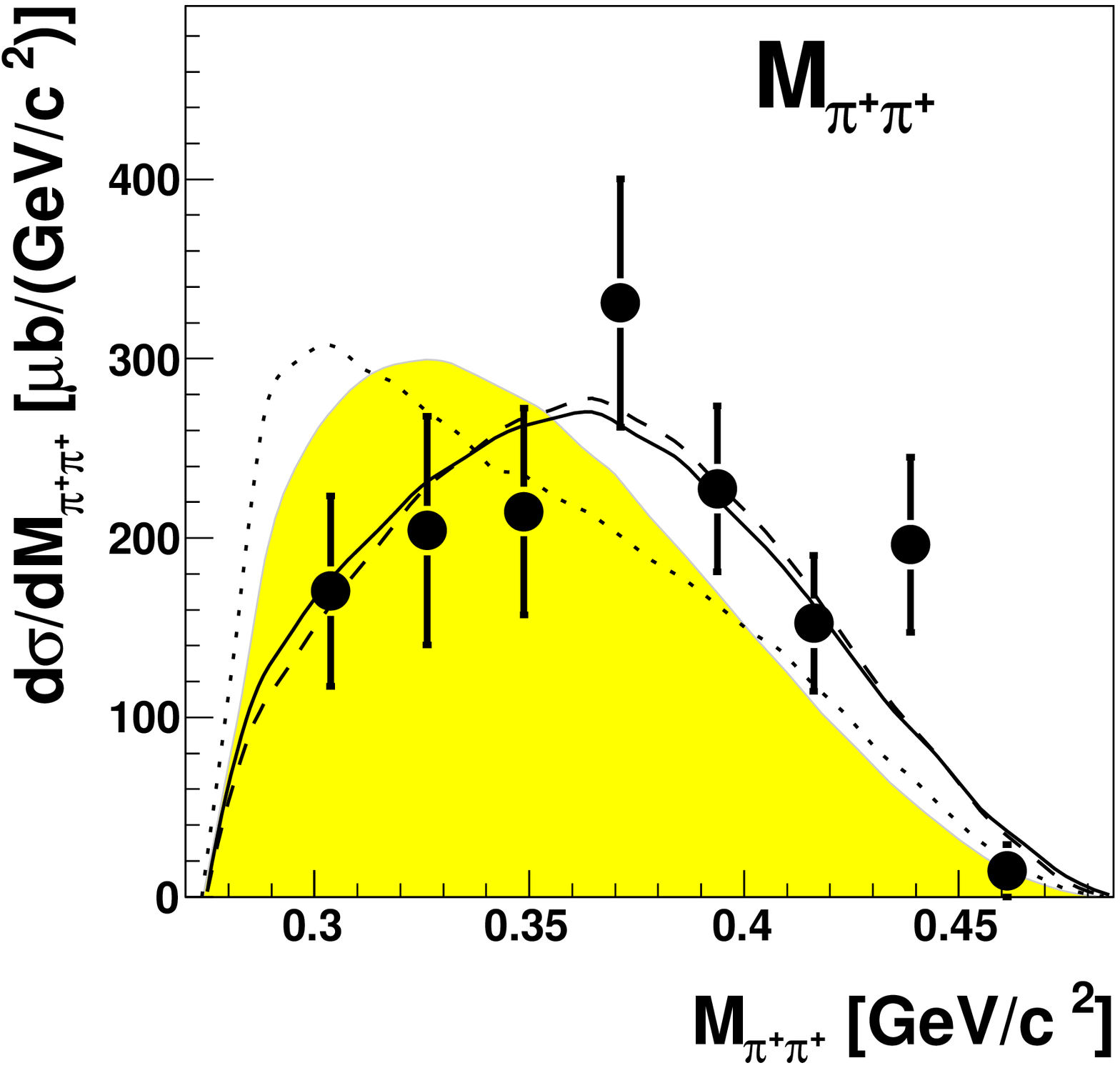}
\includegraphics[width=0.23\textwidth]{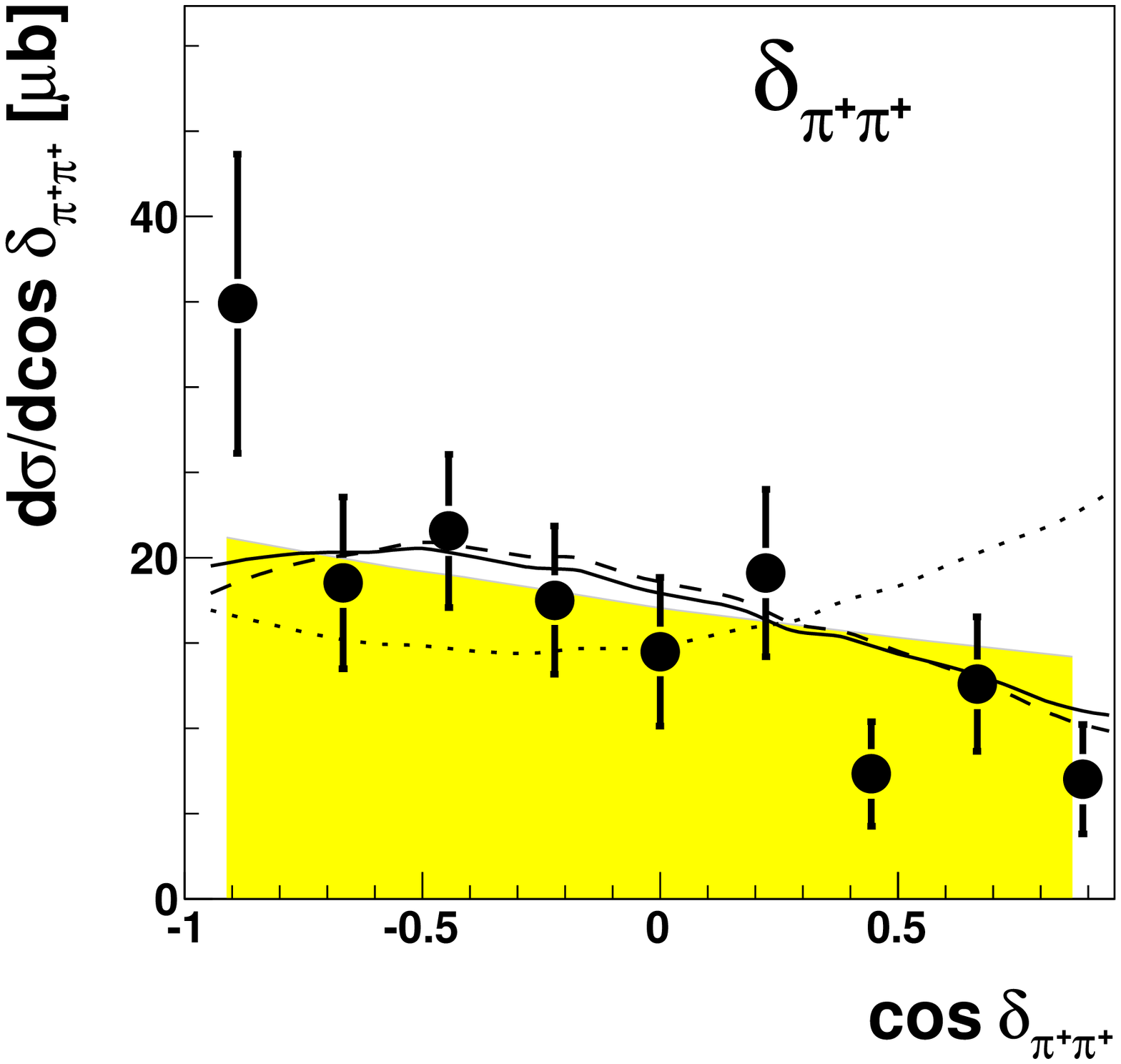}

\caption{ 
   Distribution of the $\pi^+\pi^+$ invariant mass $M_{\pi^+\pi^+}$  ({\bf
     left}) and the
   $\pi^+\pi^+$ opening angle $\delta_{\pi^+\pi^+}$  ({\bf right}) for the $pp
   \to nn\pi^+\pi^+$ reaction at $T_p$ = 1.1 GeV. Solid dots represent the
   experimental results of this work. The shaded areas denote phase space
   distributions. The dotted lines show $\Delta\Delta$ calculations as used
   for the description of the $pp \to pp\pi^0\pi^0$ reaction \cite{deldel}. The
   dashed lines are calculations of the $\Delta(1600) \to \Delta \pi$ process
   and the solid lines give the coherent sum of both processes. All
   calculations are normalized in area to the data. 
}
\label{fig1}
\end{center}
\end{figure}

\begin{figure}
\begin{center}

\includegraphics[width=0.23\textwidth]{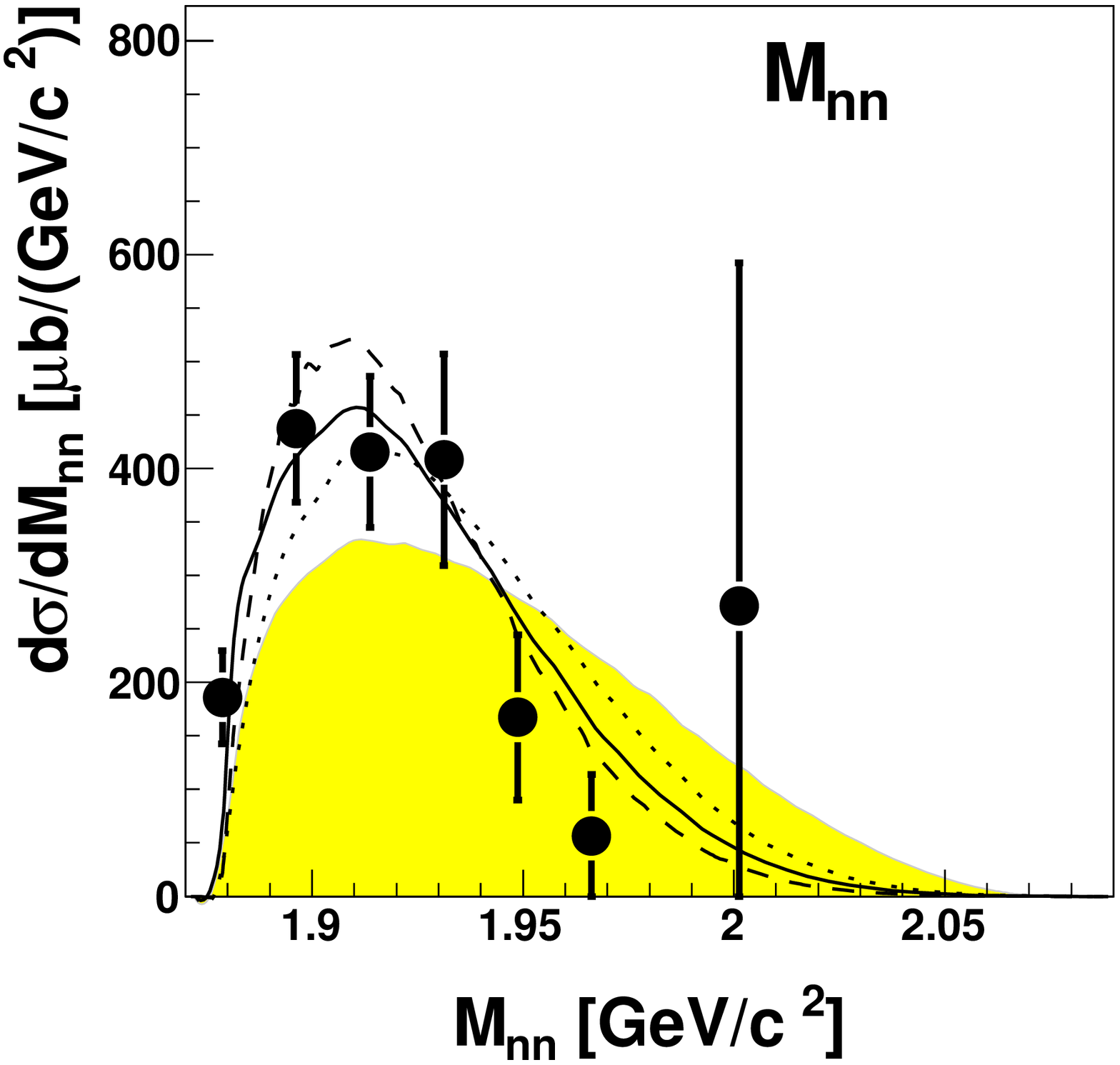}
\includegraphics[width=0.23\textwidth]{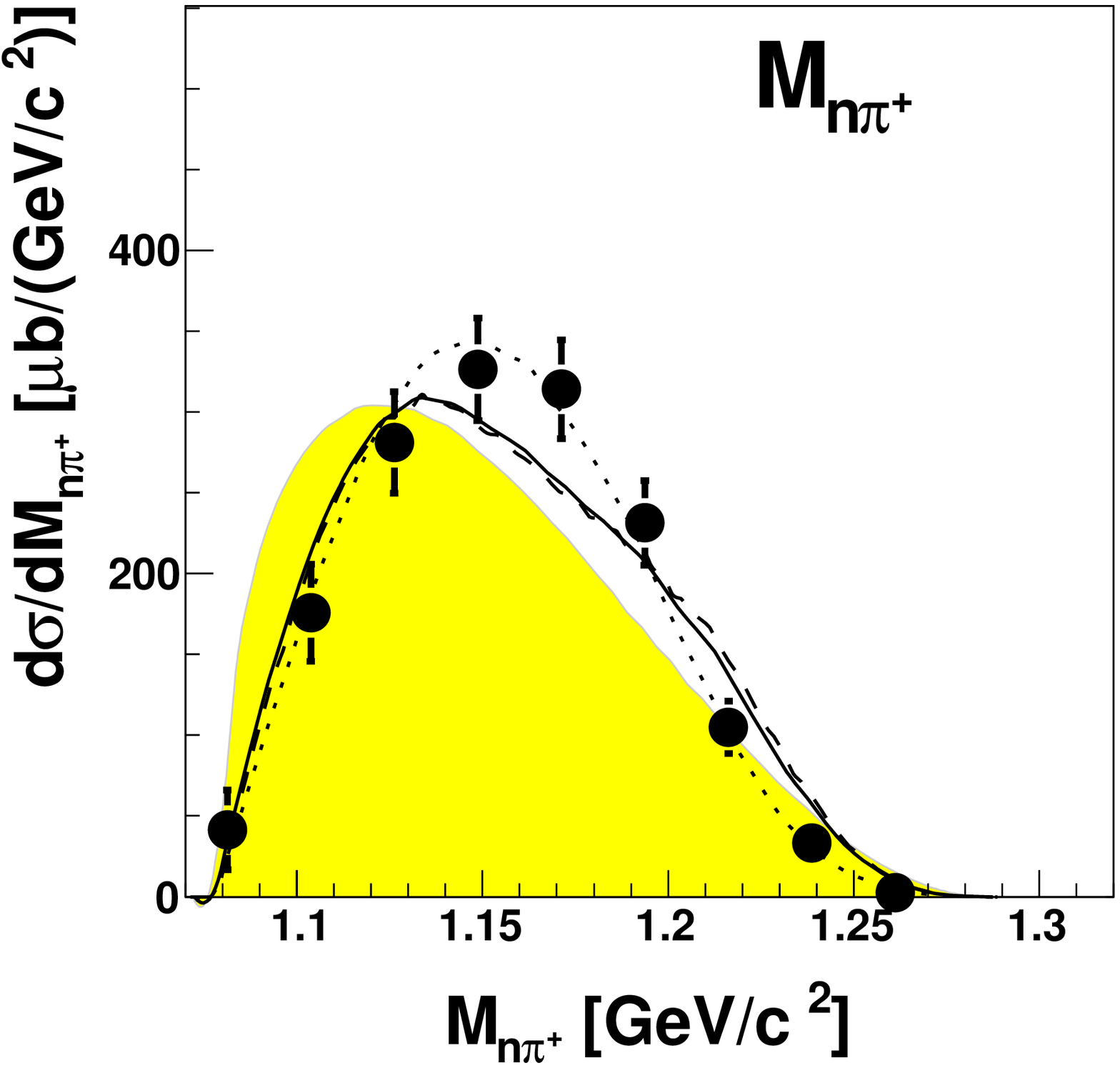}

\includegraphics[width=0.23\textwidth]{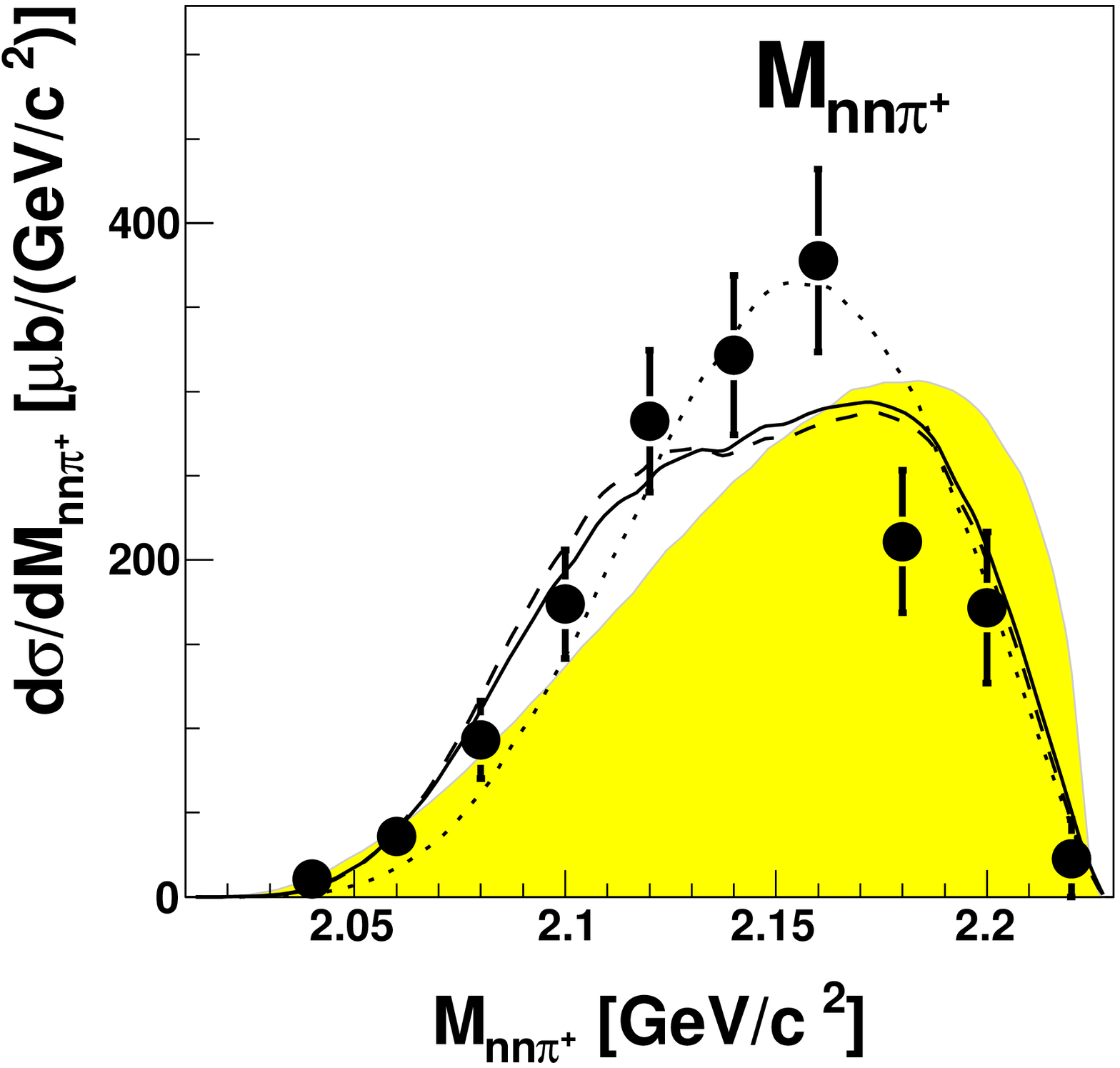}
\includegraphics[width=0.23\textwidth]{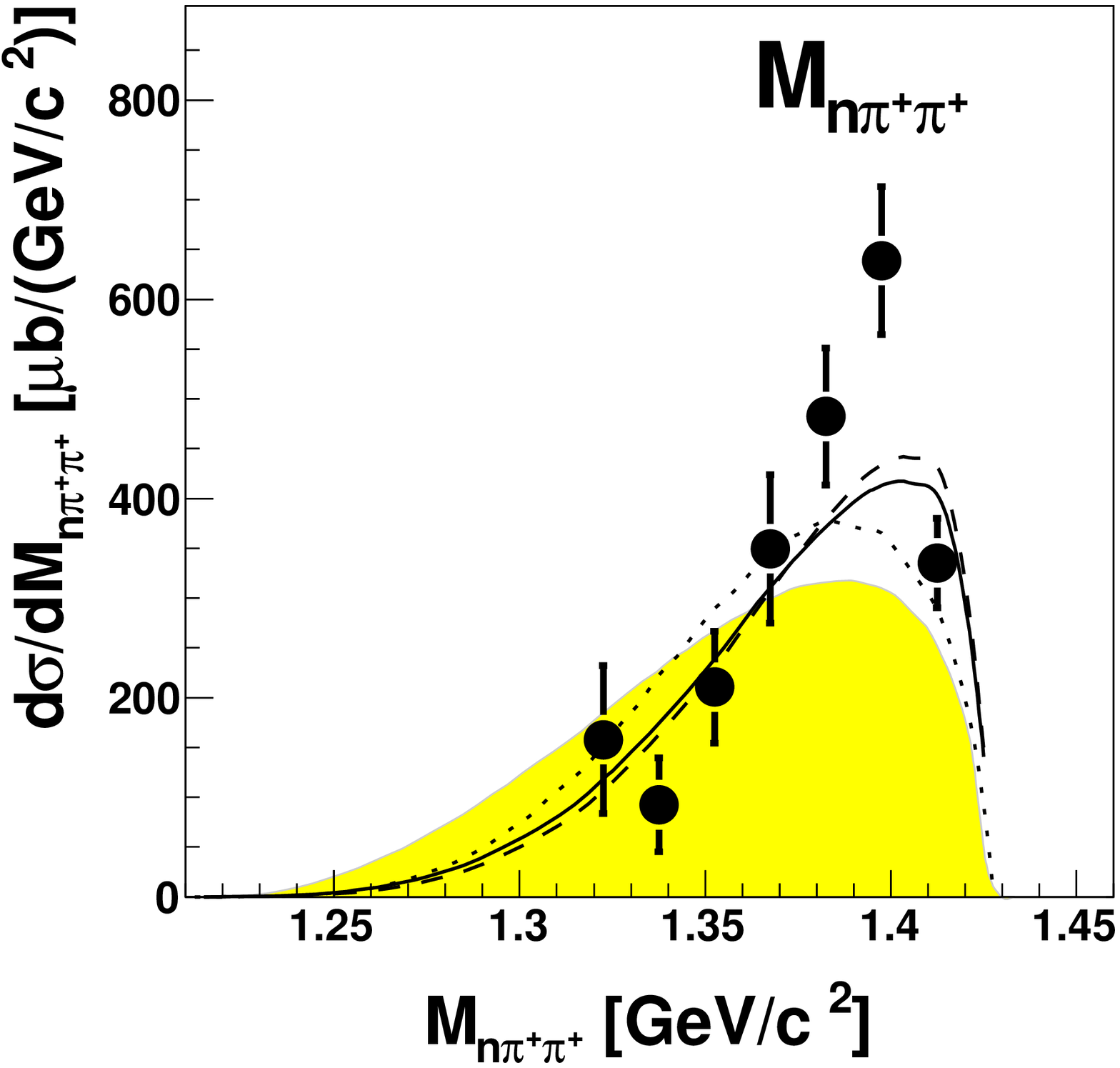}

\caption{
  Same as Fig. 2 but for the 
   distributions of the invariant masses $M_{nn}$  ({\bf
     left top}), $M_{n\pi^+}$({\bf right top}),  $M_{nn\pi^+}$  ({\bf
     left bottom}) and $M_{n\pi^+\pi^+}$({\bf right bottom}). 
}
\label{fig2}
\end{center}
\end{figure}

\begin{figure}
\begin{center}

\includegraphics[width=0.23\textwidth]{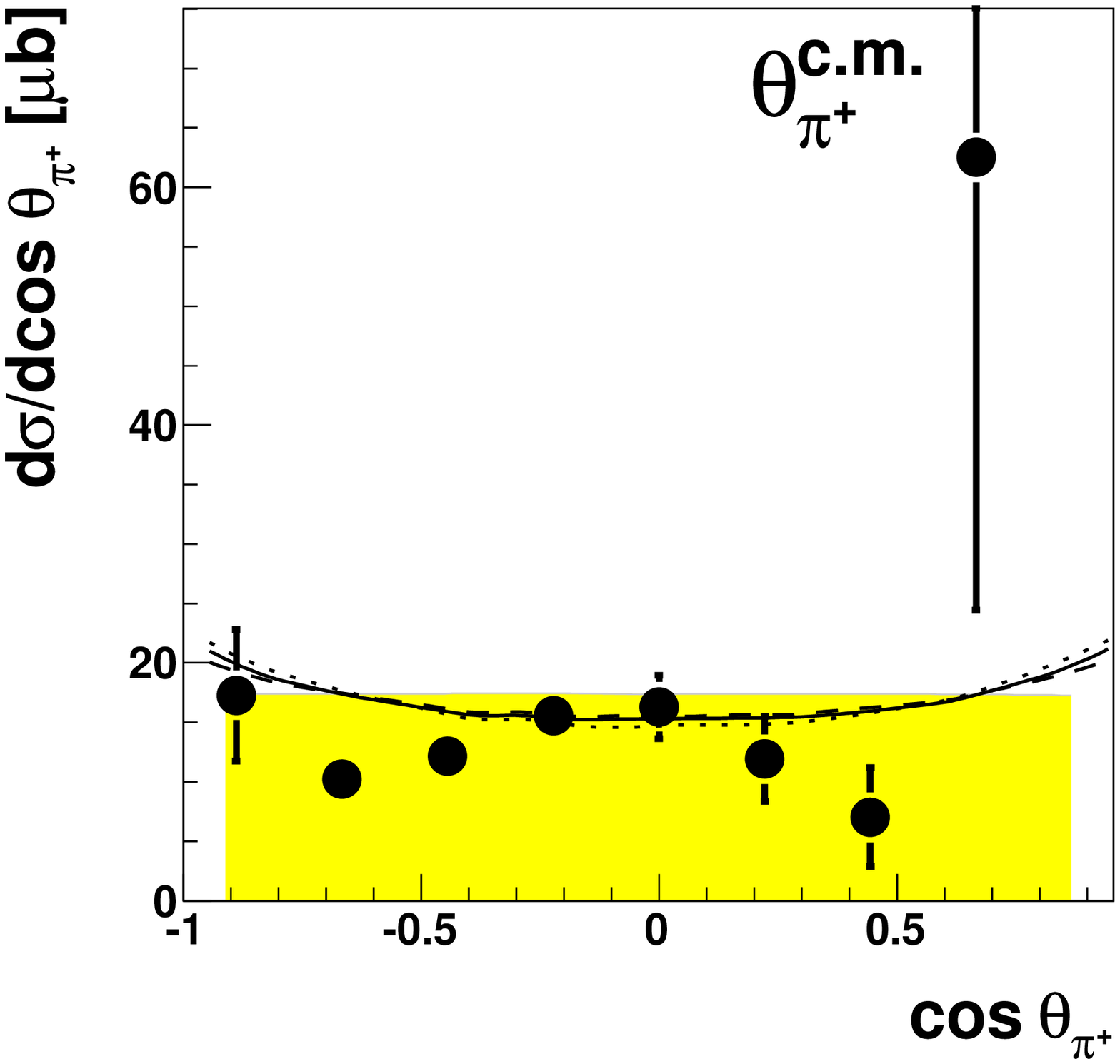}
\includegraphics[width=0.23\textwidth]{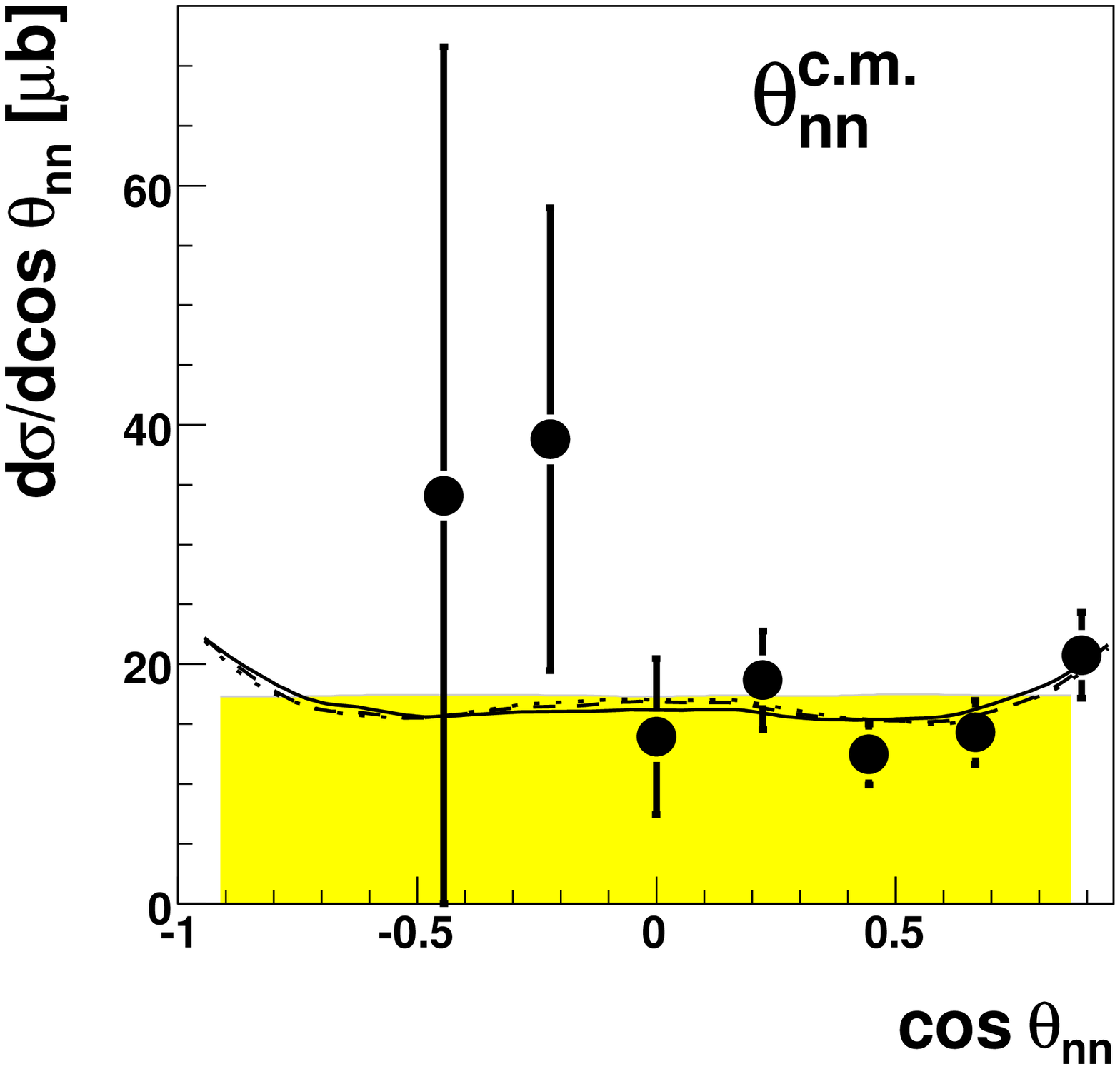}

\caption{
  Same as Fig. 2 but for the 
   distributions of the $\pi^+$ angle $\Theta_{\pi^+}^{c.m.}$  ({\bf
     left}) and of the angle 
   $\Theta_{nn}^{c.m.}$ of the $nn$ system ({\bf right}), both in the center-of-mass
   system.  
}
\label{fig3}
\end{center}
\end{figure}

The energy dependence of the total cross section is displayed in Fig. 1.
The total cross section value from this work has been published already in
Ref. \cite{tsi} in connection with the isospin decomposition of two-pion
production data.  It is in good agreement with previous bubble-chamber data
from KEK \cite{shim}, however, a factor of five larger than predicted by the
Valencia calculations \cite{alv}. This huge discrepancy was the primary reason
to introduce the excitation of a higher-lying $\Delta$  as a possible
explanation in Ref. \cite{tsi}.

The total cross section of the $pp \to pp\pi^0\pi^0$ reaction keeps rising from
threshold up to $T_p \approx$ 1 GeV, where it levels off until 1.2
GeV. Thereafter it continues steeply rising until 1.5 GeV, 
where it finally levels off again - see Figs. 1 and 3 in Ref. \cite{tsi}. As
has been demonstrated there, the low-energy structure is
due to the Roper resonance \cite{TS0}, whereas the renewed rise at higher
energies can be 
associated with the dominance of the $\Delta\Delta$ excitation
\cite{deldel}. Since the latter describes both total and differential cross
sections  for this channel quite well for $T_p >$ 1 GeV, a possibly faulty
description of the $\Delta\Delta$ process can be excluded as reason for the
failure of the theoretical prediction in the $nn\pi^+\pi^+$ channel. 

Figs. 2 - 4 show a selection of eight differential cross sections: 
the differential distributions for the invariant masses $M_{\pi^+\pi^+}$, 
$M_{nn}$, $M_{n\pi^+}$, $M_{nn\pi^+}$  and $M_{n\pi^+\pi^+}$ , the opening
angle between the two pions $\delta_{\pi^0\pi^0}$, the $\pi^+$ angle
$\Theta_{\pi^0}^{c.m.}$ as well as the angle $\Theta_{nn}^{c.m.}$ of the $nn$ system -- all
in the center-of-mass system (cms).
Note that
for a four-body reaction with unpolarized beam and target there are seven 
independent single differential distributions. The
$\delta_{\pi^+\pi^+}$ distribution in fact is correlated with the
$M_{\pi^+\pi^+}$ spectrum as discussed in some detail in
Refs. \cite{WB,JP}.

In the figures the data are compared to pure phase space distributions (shaded
areas in Figs. 2 - 4) as well as to calculations 
(dotted, dashed and solid lines in Figs. 1 - 4), which will be
discussed in the following. As a convention we show in Figs. 2 - 4 all
theoretical distributions normalized to the experimental cross section. This
is because we are interested here in the shape of the differential
distributions. For the comparison with the absolute cross section see Fig. 1.

Though the statistics of the data is limited we see that part of the
experimental differential distributions deviate significantly from phase
space, in particular the invariant mass distributions. 
We observe the $M_{\pi^+\pi^+}$ spectrum
to be broader in its distribution compared to phase space, whereas the
experimental $M_{nn}$ spectrum, which is kind of complementary to 
the $M_{\pi^+\pi^+}$ spectrum, is substantially narrower than
phase space. This behavior signals the excitation of a system which requires a
large internal energy: in order to reach a maximum possible
excitation energy of this system, the two involved neutrons must have minimal
relative kinetic energy.

Next we confront the data with theoretical predictions \cite{alv} of the
Valencia group and subsequent modifications of the original calculations. In
Ref. \cite{alv} three processes have been considered to feed the $pp \to
nn\pi^+\pi^+$ reaction: 
\begin{itemize}
\item the $\Delta\Delta$ process, which is considered to be
the leading process for $T_p >$ 1 GeV, 
\item the contribution from nonresonant chiral terms according to graphs (1) -
  (3) in Ref. \cite{alv}, which shows a phase space like behavior and is
  expected to be the leading contribution for $T_p <$ 1 GeV and finally 
\item the contribution from the excitation of the Roper resonance with
  subsequent single-pion decay and associated nonresonant emission of a second
  pion, graphs (6) - (7) in Ref. \cite{alv}. This contribution is the
  smallest one in the calculations of Ref. \cite{alv}. Accounting for the
    finding in Ref. \cite{TS0} that the Roper contribution is largely
    overestimated for $T_p >$ 1 GeV in  Ref. \cite{alv},
    this contribution should be insignificant for the data discussed here. We
    note in passing that in principle also a double Roper excitation,
    i.e. excitation of the Roper resonance in each of the participating
    nucleons could feed the $pp \to nn\pi^+\pi^+$ reaction, however, such a
    process must be even much rarer than the single Roper process discussed
    here.   
\end{itemize}

In Ref. \cite{deldel} we have demonstrated that the differential data for
the $pp \to pp\pi^0\pi^0$ channel in 
the region of the $\Delta\Delta$ excitation are not well described by
Ref. \cite{alv}, which assumes $\rho$ exchange to be dominating for this
process. However, with $\pi$ exchange as the leading process good agreement
with the data is found. This is in accordance with the finding of
Ref. \cite{xu} that $\rho$ exchange is of minor importance.  

The dotted lines in Figs. 2 - 4 show the predictions for the
$\Delta\Delta$ process calculated according to Ref. \cite{deldel}, which though
renormalized in area give large deviations from the measured distributions - in
particular in the $M_{\pi^+\pi^+}$ and $\delta_{\pi^+\pi^+}$ spectra. For the
$M_{\pi^+\pi^+}$ spectrum these calculations predict a low-mass
enhancement, which is absent in 
the data. With regard to the $\pi^+\pi^+$ opening angle $\delta_{\pi^+\pi^+}$
these calculations predict preferential parallel and antiparallel emissions
of the two pions, which again is not supported by the data. 


Next we calculate the excitation of the $\Delta(1600)$ and their subsequent
decay $\Delta(1600)^{++} \to \Delta^+\pi^+ \to n\pi^+\pi^+$ by modifying graph
(9) of Ref. \cite{alv} accordingly. This calculation is shown by the
dashed lines. They differ from the $\Delta\Delta$ results in particular in the
$M_{\pi^+\pi^+}$ and $\delta_{\pi^+\pi^+}$ spectra reaching there good 
agreement with the data. Finally we superimpose the $\Delta(1600)$ and 
$\Delta\Delta$ processes destructively as required by the isospin decomposition
results \cite{tsi}. This is shown by the solid lines, which actually are very
close to 
the dashed ones, since the  $\Delta(1600)$ process is the dominating
process. The strength of this process has been adjusted to fit the total cross
section data.

Note that since both processes are connected with a double $p$-wave emission
of the pion pair, the angular distributions for pions and neutrons are very
similar. Hence these angular distributions cannot discriminate between both
processes.

Summarizing we have presented first exclusive 
measurements of the $pp \to nn\pi^+\pi^+$ reaction. The data show evidence for
excitation and decay of a higher-lying $\Delta$ resonance, possibly the
$\Delta(1600)$. The description of the differential data is consistent with
that of the total cross section and with the description of the data in other
channels as required by the isospin decomposition.


We acknowledge valuable discussions with L. Alvarez-Ruso, C. Hanhart, E. Oset
and C. Wilkin on this issue. We are particularly indebted to L. Alvarez-Ruso
for using his code.  
This work has been supported by BMBF
(06TU9193), Forschungszentrum J\"ulich (COSY-FFE) and  
DFG (Europ. Graduiertenkolleg 683). 

\end{document}